\documentclass{article}

\usepackage{natbib}
\usepackage{graphicx}
\usepackage{arxiv}

\usepackage{amssymb, amsmath, dsfont}
\usepackage{appendix}

\usepackage{multirow}
\usepackage[lofdepth,lotdepth]{subfig}

\newcommand{\bW}{\mathbf{W}}
\newcommand{\bH}{\mathbf{H}}
\newcommand{\bh}{\mathbf{h}}
\newcommand{\bX}{\mathbf{X}}
\newcommand{\bx}{\mathbf{x}}

\newcommand{\bw}{\mathbf{w}}

\DeclareMathOperator{\trace}{Trace}

\title{PIntMF: Penalized Integrative Matrix Factorization method for Multi-omics data}
\author{Morgane PIERRE-JEAN\\ Universit\'e de Paris-Saclay, \\Centre National de Recherche en G\'enomique Humaine, CEA, Evry, France, \\
\texttt{mpierrejean.pro@gmail.com} 
\And Florence MAUGER\\ Universit\'e de Paris-Saclay, \\Centre National de Recherche en G\'enomique Humaine, CEA, Evry, France\\
\And Jean-Fran\c cois DELEUZE\\
Universit\'e de Paris-Saclay, \\Centre National de Recherche en G\'enomique Humaine, CEA, Evry, France\\
\And Edith LE FLOCH\\ Universit\'e de Paris-Saclay, \\Centre National de Recherche en G\'enomique Humaine, CEA, Evry, France}

\begin{document}


\maketitle

\begin{abstract}
It is more and more common to explore the genome at diverse levels and not only at a single omic level. Through integrative statistical methods, omics data have the power to reveal new biological processes, potential biomarkers, and subgroups of a cohort.
The matrix factorization (MF) is a unsupervised statistical method that allows giving a clustering of individuals, but also revealing relevant omic variables from the various blocks. 
 Here, we present PIntMF (Penalized Integrative Matrix Factorization), a model of MF with sparsity, positivity and equality constraints.
To induce sparsity in the model, we use a classical Lasso penalization on variable and individual matrices. For the matrix of samples, sparsity helps for the clustering, and normalization (matching an equality constraint) of inferred coefficients is added for a better interpretation.
Besides, we add an automatic tuning of the sparsity parameters using the famous \texttt{glmnet} package. We also proposed three criteria to help the user to choose the number of latent variables. PIntMF was compared to other state-of-the-art integrative methods including feature selection techniques in both synthetic and real data. PIntMF succeeds in finding relevant clusters as well as variables in two types of simulated data (correlated and uncorrelated). Then, PIntMF was applied to two real datasets (Diet and cancer), and it reveals interpretable clusters linked to available clinical data. Our method outperforms the existing ones on two criteria (clustering and variable selection). We show that PIntMF is an easy, fast, and powerful tool to extract patterns and cluster samples from multi-omics data.\\
\end{abstract}

\section{Introduction}
The improvement of high-throughput biological technologies enables the production of various omics data such as genomic, transcriptomic, epigenomic, proteomic, and metabolomic data \citep{ritchie2015methods,yugi2016trans}. The generation of these data allows investigating biological processes in cancer or complex diseases. For example,  The Cancer Genome Atlas (TCGA \citep{cancer2012comprehensive}) has already produced numerous omics data for a set of 32 cancer types \citep{vasaikar2017linkedomics}. Recently, other multi-omics studies on complex diseases and single-cell data are also emergent \citep{rowlands2014multi,bock2016multi,yang2020multitissue}.

However, integrating omics data addresses several statistical challenges, such as dealing with a large number of variables, few samples, and data heterogeneity \citep{bersanelli2016methods}. Indeed, the statistical distributions of omics data are very heterogeneous. For instance, mutations can be modeled by a binary distribution, while RNAseq data can be modeled by a Negative Binomial distribution and metabolomic data by a Gaussian distribution. Besides, the omic block sizes could vary from one hundred to one billion variables. Furthermore, collecting several omics for a single sample could be difficult due to the cost and access to the biological material.

 To identify potential biomarkers and new classifications in complex diseases, since the last decade, unsupervised integrative methods have been developed to analyze the multi-omics datasets \citep{tini2017multi,huang2017more,chauvel2019evaluation, pierre2019clustering, Cantini2020}. Blocks of omics data can be seen as matrices, and relevant information can be extracted using dimension reduction methods, particularly, matrix factorization (MF) methods \citep{sastry2020matrix} and canonical correlation analysis (CCA) \citep{tenenhaus2011regularized}. 
 
CCA methods are used to integrate multi-omics data and aim to maximize the correlation between omics under constraints \citep{tenenhaus2011regularized, tenenhaus2014variable, rodosthenous2020integrating}.

Then, MF techniques infer two matrices when applied to a single omic data: the first one describes the structure between variables (e.g., genes, probes, regions) and the second one describes the structure between samples.

One famous MF method is the Non-Negative Matrix Factorization (NMF, \citep{lee1999learning}). This method implements non-negativity constraints on the two inferred matrices.
NMF provides a way to explain the structure of data by providing variable profiles (dictionary for each dimension). Besides, NMF enables a classification of the samples thanks to the second matrix. The NMF is a commonly applied method used for a single omic block to identify disease subtypes in gene expression data  \citep{burstein2015comprehensive} or recently, in DNA methylation data \citep{reilly2019dna}.

 More recently, extensions of MF have been developed to perform integrative analysis \citep{mo2013pattern,chalise2014integrative,chen2018discovery}. MF extensions need to infer more than two matrices: one matrix for each omic block is computed and one matrix for samples.

Matrix Factorization showed that it is a powerful technique to integrate heterogeneous data \citep{chauvel2019evaluation, pierre2019clustering, Cantini2020}. In our article, we propose a Penalized Integrative Matrix Factorization method called PIntMF, to discover new patterns and a new classification of a cohort. First, to add sparsity on the first inferred matrix (corresponding to the variable blocks), we used a common regularization technique: the Least Absolute Shrinkage and Selection Operator (LASSO \citep{tibshirani1996regression}). Then, sparsity, non-negativity and equality constraints are added to the second matrix (corresponding to the samples) to help for the interpretability of the clustering.

Besides, we propose criteria to choose the number of latent variables and to properly initialize the algorithm.

The performance of this new unsupervised model was evaluated on both simulated and real data. We applied PIntMF on a simulated framework introduced by our group in  \citep{pierre2019clustering} but also on a simulated framework from \citep{chung2019multi}. We compared our method to several existing unsupervised methods that perform both variable selection and clustering: intNMF  \citep{chalise2017integrative}, SGCCA \citep{tenenhaus2014variable}, MoCluster \citep{meng2015mocluster}, CIMLR \citep{ramazzotti2018multi}, and iClusterPlus \citep{iclusterPlus}. 
Then, we applied the model on a murine liver dataset \citep{williams2016systems} and glioblastoma cancer data from TCGA already used in \citep{shen2012integrative}.

\enlargethispage{-65.1pt}

\section{Method}
\subsection{Model description}

In the following, $\mathbf{A}$ denotes a matrix, $\mathbf{a}$ a vector and $a$ a scalar. We consider $K$ matrices $\mathbf{X}_1,  \dots \mathbf{X}_K$ as the input of each method. Each matrix $\mathbf{X}_k$ is of size $n\times J_k$ ($n$ is the number of samples and $J_k$ the number of variables for the block $k$.
In this article, we propose a model based on the matrix factorization method i.e.:

\begin{equation}
\label{eq:MF}
\bX^k \approx\bW\bH^k
\end{equation}

where $\bW$ denotes  a common basis matrix and $\bH^k$ a specific coefficient matrix associated with the block $k$. $\bW$ is of size $n\times P$ and $\bH^k$ is of size $P\times J_k$. Therefore, the variable $P$ is the number of latent variables in the model.

To ensure identifiability and improve interpretation of the model, non-negativity and sparsity constraints are imposed on $\bW$ (as in intNMF model described in \citep{chalise2017integrative}). $\bW$ will be used to cluster samples simultaneously across the $K$ omics blocks. On $\bH^k$, a sparsity constraint is imposed to perform variable selection simultaneously to the clustering of samples.
The model \ref{eq:MF} can be extended to the following optimization problem:

\begin{eqnarray}
    \label{eq:my-model}
    \min_{\bW, \bH_1,\dots, \bH^k}\sum_{k=1}^K \Vert\bX^k -\bW\bH^k\Vert^2_F+ 
    \lambda_k\Vert\bH^k\Vert_1+ \\\nonumber\sum_{i=1}^n\mu_i\Vert\bw_{i\bullet}\Vert_1 \\\nonumber
    \quad\text{s.t.} \quad\bW \geq0
\end{eqnarray}
 where $\Vert\bH^k\Vert_1= \sum_{p=1}^P\sum_{j=1}^{J_k}\vert h^k_{pj}\vert$.
 
\subsection{Solving equation}

The optimization problem \ref{eq:my-model} is not convex on $\bW, \bH_1,\dots, \bH^k$, but is convex separately on each matrix.
Consequently, it can be solved alternatively on $\bW, \bH_1,\dots, \bH^k$ until convergence.

\subsubsection{Solve on $\bW$}

In this step, $\bH^k$ is fixed and the problem \ref{eq:solvew} is solved on $\bW$.
\begin{equation}
    \label{eq:solvew}
    \min_{\bW}\sum_{k=1}^K \Vert\bX^k -\bW\bH^k\Vert^2_F +\sum_{i=1}^n\mu_i\Vert\bw_{i\bullet}\Vert_1 \quad\text{st.} \quad\bW \geq0
\end{equation}

All individuals are independent for the weights $\bW$ when $\bH^k$ are fixed. 
The problem for an individual $i$ can be written as follows:

\begin{equation}
    \label{eq:solvew_2}
    \min_{\bw_{i\bullet}}\sum_{k=1}^K \Vert\bx_{i\bullet}^k -\bw_{i\bullet}\bH^k\Vert^2 +\mu_i\Vert\bw_{i\bullet}\Vert_1 \quad\text{st.} \quad\bw_{i\bullet} \geq0
\end{equation}

Equation \ref{eq:solvew_2} is equivalent to 
\begin{equation}
    \label{eq:solvew_3}
    \min_{\bw_{i\bullet}}\sum_{k=1}^K\sum_{j=1}^{J_k} (x_{ij}^k -\bw_{i\bullet}\bh_{\bullet j}^k)^2 +\mu_i\Vert\bw_{i\bullet}\Vert_1 \quad\text{st.} \quad\bw_{i\bullet} \geq0
\end{equation}

The optimization problem described by \ref{eq:solvew_3} is a classical lasso problem with a positivity constraint. It can be easily and fastly solved by \texttt{glmnet R} package \citep{glmnet}. 

\subsubsection{Solve on $\bH^k$}

When $\bW$ is fixed, each $\bH^k$ can be solved independently. In this section, to be more readable, the index $k$ is removed from the equations.

\begin{equation}
    \label{eq:H-solve}
    \min_{\bH}Q(\bH)= \min_{\bH}\Vert\bX -\bW\bH\Vert^2_F+ \lambda\sum_{p=1}^P\sum_{j=1}^{J}\vert h_{pj}\vert
\end{equation}

\vspace{0.5cm}

$\begin{array}{rl}
    Q(\bH) =&\trace\left\lbrace(\bX-\bW\bH) (\bX-\bW\bH)^T\right\rbrace+\\
        \vspace{0.5cm}
&\lambda\sum_{p=1}^P\sum_{j=1}^{J}\vert h_{pj}\vert \\
     =&  vec(\bX-\bW\bH)^Tvec(\bX-\bW\bH)+\\&\lambda\sum_{p=1}^P\sum_{j=1}^{J}\vert h_{pj}\vert
\end{array}$

\vspace{0.5cm}

We denote $\bh=vec(\bH) = \left(\begin{array}{c}
     \bH_{11}  \\
     \vdots \\
     \bH_{P1}\\
     \vdots\\
     \bH_{1J}\\
     \vdots\\
     \bH_{PJ}\\
\end{array}\right)$  and $\bx=vec(\bX)=\left(\begin{array}{c}
     \bX_{11}  \\
     \vdots \\
     \bX_{n1}\\
     \vdots\\
     \bX_{1J}\\
     \vdots\\
     \bX_{nJ}\\
\end{array}\right)$.

\vspace{0.5cm}

$\begin{array}{rl}
    Q(\bH) =&(\bx-vec(\bW\bH))^T(\bx-vec(\bW\bH))+\lambda\Vert \bh\Vert_1\\
    =&(\bx-(\mathbb{I}_J	\otimes\bW)vec(\bH))^T(\bx-(\mathbb{I}_J	\otimes\bW)vec(\bH))\\
    &+\lambda\Vert \bh\Vert_1\\
     =&(\bx-\tilde{\bW}\bh)^T(\bx-\tilde{\bW}\bh)+\lambda\Vert \bh\Vert_1
\end{array}$

where $\mathbb{I}_J$ is the identity matrix of size $J$ and  $\tilde{\bW}=\mathbb{I}_J	\otimes\bW$

We can reformulate the problem as follows:\\
$$\begin{array}{rl}
    Q(\bH) =&\Vert\bx-\tilde{\bW}\bh\Vert^2+\lambda\Vert \bh\Vert_1
\end{array}$$

$\lambda$ will be optimized for each block $k=1, \dots, K$.

As for $\bW$, we used the \texttt{glmnet} package to solve this problem.

\subsubsection{Normalization}

We would like to consider $\bW$ as a weight matrix. To avoid problems of convergence or non-identifiability, the normalization by the sum of weights for each row of $\bW$ is added after computing the matrix, i.e. each row is divided by its sum after each step:
\begin{equation}
    \bw_{i\bullet}= \frac{\bw_{i\bullet}}{\sum_{p=1}^P\bw_{ip}}
\end{equation}

\subsection{Stopping criteria}

The stopping criterion of the model is determined by the convergence of the matrix $\bW$. The stability of the similarity of matrix $\bW$ between two iterations means that the model has converged therefore we stop the algorithm. The similarity between $\bW^{t-1}$ and $\bW^{t}$ is measured  with the ARI. The users have also the possibility to define a maximum number of iterations to limit the time of the algorithm.

\subsection{Automatic tuning of sparsity parameters}
For each block $\bX^k$, we need to calibrate the sparsity parameter $\lambda_k$ and  $\mu_i$. The main advantage of \texttt{glmnet} package is the speed (see Supplementary Materials Fig. S9). Besides, \texttt{glmnet} implements a cross validation technique to choose the best $\lambda$ or $\mu$.
PIntMF takes advantage of \texttt{glmnet} to calibrate the penalty on each block. 
Therefore the only parameter that the user needs to tune is the number of latent variables $P$.

\subsection{Clustering}

In this article, all clusterings are obtained by applying a hierarchical clustering with the ward distance \citep{ward1963hierarchical} on matrix $\bW$. For the optimal number of clusters, $P$ is chosen.

\subsection{Criteria to choose the best model}

In this section, we present three different criteria to choose the appropriate number of latent variables ($P$). 

\subsubsection{Mean square error}
The number of latent variables can be optimized by looking at the curve of the Mean Square Error (MSE).
In this context, the mean square error (MSE) for each dataset $k$ is defined by:
\begin{equation}
    MSE^k_P = \frac{\Vert \bX^k-\bW\bH^k\Vert^2_F}{n\times J_k}
\end{equation}
Then, the total MSE is then defined by averaging the different $MSE^k_P$:
\begin{equation}
      MSE_P = \sum_k MSE^k_P/K
\end{equation}

\subsubsection{Percentage of variation explained (PVE)}

To measure the performance of the method, we computed the Percentage of Variation Explained \citep{nowak2011fused} defined by the following formula:

\begin{equation}
    PVE(\bW, \bH^k)= 1-\frac{\Vert \bX^k-\bW\bH^k\Vert^2_F}{\Vert \bX^k-\Bar{\bX}^k\mathbf{1}_{Jk}\Vert^2_F}
\end{equation}
where $\Bar{\bX}^k$ is a vector containing the average profile of each individual:

$\Bar{\bX}^k_i= \frac{\sum_j x_{ij}}{J_k}$, and $\mathbf{1}_{Jk}=(1,\dots,1)$ is a row-vector of size $J_k$.

Then, we computed the global PVE as the mean of the PVE on the $K$ blocks i.e.:

\begin{equation}
    PVE= \frac{1}{K}\sum_{k=1}^KPVE(\bW, \bH^k)
\end{equation}

\subsubsection{Cophenetic distance}
We were inspired by \citep{gaujoux2010flexible} for the last criterion.


We want to assess if the distances in the tree  (after hierarchical clustering on $\bW$) reflect the original distances accurately.

One way is to compute the correlation between the cophenetic distances and the original distance data generated by the dist() function on $\bW$ \citep{sokal1962comparison}. The clustering is valid, if the correlation between the two quantities is high. Note that we use the cophenetic function defined by \citep{sneath1973numerical}.

The cophenetic correlation usually decreases with the increase of $P$ values. 
\cite{brunet2004metagenes}  suggested choosing the smallest value of $P$ for which this coefficient starts decreasing.

\section{Performance criteria}

Two criteria are used to assess the performance of our method and to compare it with others.

\subsection{Adjusted Rand Index (ARI)}

 \label{sec:ari}
On a simulated dataset and on well known real datasets, it is possible to compute the similarity between the true and the inferred classifications. We use the Adjusted Rand Index as a criterion to evaluate the performance of our method. The Adjusted Rand Index \citep{rand1971objective} is equal to one when the two classifications that are compared are totally similar and zero or even negative if the classifications are completely different.

\subsection{Area under the ROC curve (AUROC)}
 \label{sec:roc}
On a simulated dataset, the variables that drive the subgroups are known, and it is easy to compute false-positive and true-positive rates. First, variables are ordered by their standard deviation (from the highest to the lowest) computed on the $\bH$ matrix to highlight the largest differences between the $P$ components and therefore the most contributory to the clusters. To summarize the information of these two quantities, we compute the area under the TPR-FPR curve (AUROC). An AUROC equal to one means that the method selects the variables with no error. An AUROC under 0.50 means that false-positive variables are selected before the true positive ones.

\section{Results}

\subsection{Optimization of the algorithm}
\subsubsection{Initialization}
Often in NMF algorithms \citep{lee1999learning}, the matrices are initialized by non-negative random values. We assess four kinds of initialization for PIntMF (hierarchical clustering, random, Similarity Network Fusion and Singular Values Decomposition).

The best initialization is based on the SNF algorithm \citep{wang2014similarity} (Fig. S1). This initialization has the advantage to take into account simultaneously the $K$ blocks of the analysis.

Therefore, for all the following analyses, SNF initialization was used.

\subsubsection{Computing optimization of $\bH$}

Several algorithms to solve the Lasso problem on $\bH^k$ were tested. \texttt{glmnet} is the fastest package among them (Supplementary materials Fig. S9).

\subsection{Performance on simulated datasets}

We assess the performance of PIntMF in two simulated frameworks described below.

\subsubsection{Simulations on independent datasets (non-correlated blocks)}
\label{sec:sim}
The performance of PIntMF to cluster samples and to select relevant variables was evaluated on simulated data described in \citep{pierre2019clustering}. The framework of these simulations is composed of three blocks with three different types of distribution (Binary, Beta-like, and Gaussian) to simulate the heterogeneity of the integrative omics data studies. 
Indeed, a binary distribution could match a mutation (equal to 1 if the gene is mutated and 0 otherwise); a Beta-like distribution could match DNA methylation data, and a Gaussian distribution could match gene expression values.

Four unbalanced groups (composed of 25, 20, 5, and 10 individuals) have been simulated (Benchmarks 1 to 5). Datasets with 2, 3, and 4 balanced groups have also been simulated (Benchmarks 6 to 8). Each benchmark is simulated 50 times. 

PIntMF was compared to several integrative unsupervised methods \citep{pierre2019clustering} that perform both clustering and variable selection namely: intNMF \citep{chalise2014integrative}, SGCCA \citep{tenenhaus2014variable}, MoCluster \citep{meng2015mocluster}, iClusterPlus \citep{mo2013pattern}, and CIMLR \citep{ramazzotti2018multi}.

Clustering performance was evaluated using the Adjusted Rand Index on simulated data (see section \ref{sec:ari}).

\begin{figure*}[ht]
    \centering
\includegraphics[width=0.99\textwidth]{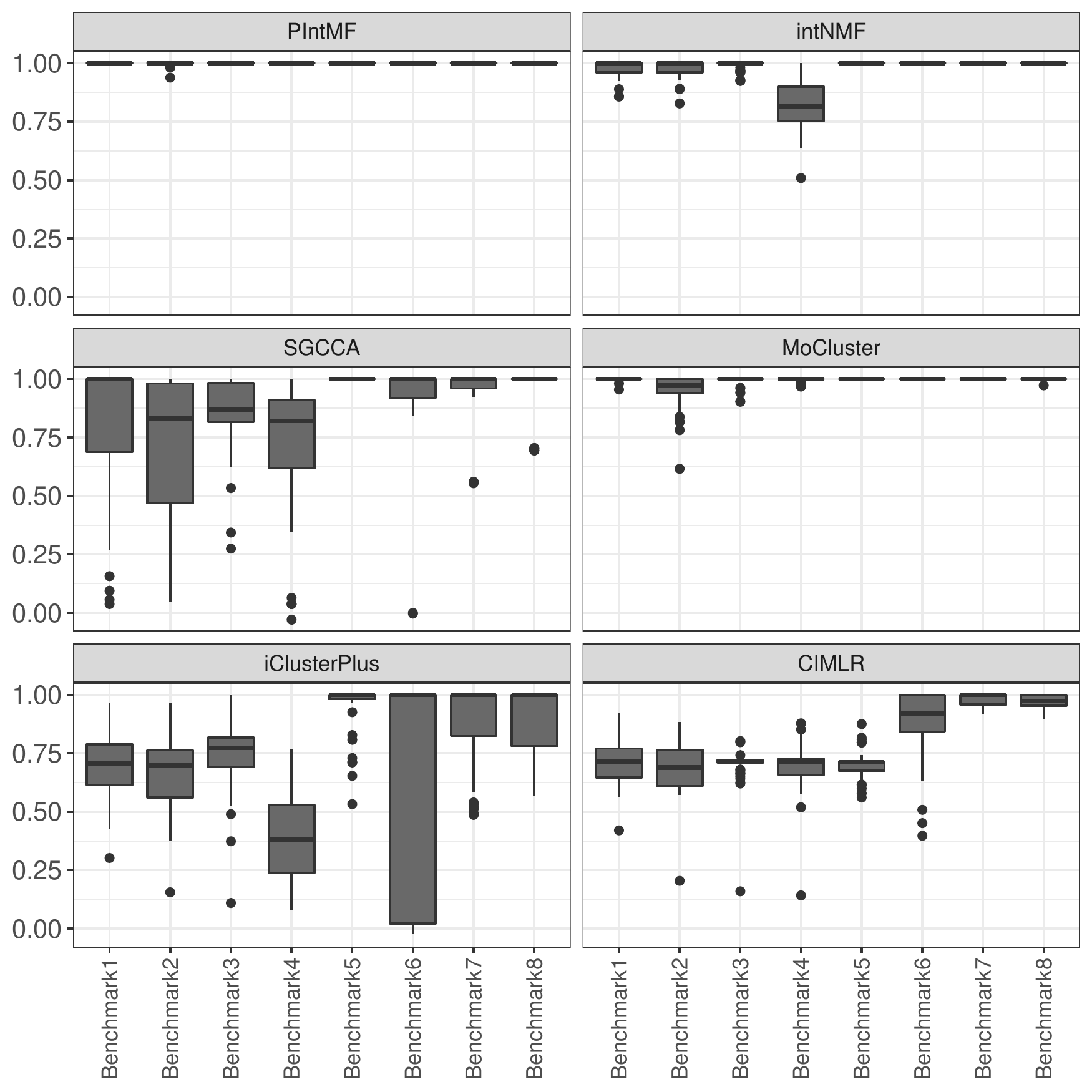}
    \caption{Adjusted Rand Index of PIntMF, intNMF, SGCCA, MoCluster, iClusterPlus, and CIMLR methods on simulated datasets. B1 :Referrence, B2: More Gaussian noise, B3: More Gaussian noise and more Binary noise, B4: More Beta noise and more Binary noise, B5: More Relevant variables, B6: 2 balanced groups, B7: 3 balanced groups, B8: 4 balanced groups}
    \label{fig:Clust-sim}
\end{figure*}

On the eight simulated benchmarks with various levels of signal to noise ratio, PIntMF and MoCluster outperform the other methods with an ARI equal to 1 in most cases (Fig. \ref{fig:Clust-sim}).

The performance of variable selection is assessed using the area under ROC curves (AUROC) after computing False Positive Rates (FPR) and True Positive Rates (TPR) (see section \ref{sec:roc}). The computation of the AUROC shows that PIntMF performs as well as MoCluster on the three types of data (Table S1 in Supplementary Materials). Indeed, PIntMF reaches either the first or the second-best AUROC for these simulations. Besides, the lowest AUROC is equal to 0.88 which means that the method is both sensitive and specific.

\subsubsection{Simulation based on real data (correlated blocks)}
\label{sec:sim-real}

We evaluate the performance of PIntMF on a simulated framework based on cancer real data and developed by \citep{chung2019multi}. Indeed, the previous framework does not simulate any correlation between omics blocks. 

OmicsSIMLA is a simulation tool for generating multi-omics data with disease status. This tool simulates CpGs with methylation proportions, RNA-seq read counts and normalized protein expression levels. Here, we simulated 50 datasets containing 50 cases (i.e., short-term survival) and 50 controls (i.e. long-term survival), and three omics blocks (RNAseq, DNA methylation, and proteins). We try to recover the two groups but also the different features that drive overall survival by using DNA methylation, expression, and protein data. For two of the three blocks (expression and DNA methylation), the variables differentially expressed or methylated between the two groups are known.

The simulated data are described in Supplementary Materials (Section 5).

In these simulations, we also compare the performance of PIntMF to other methods in terms of clustering and variable selection.
First, CIMLR does not give any results on these simulations (the algorithm does not converge). 
For all the other methods, the ARI is equal to 1 (maximum value) for all 50 datasets. 

Then, we compare the variable selection performance of PIntMF, intNMF, iClusterPlus, MoCluster, and SGCCA by computing the AUROC on expression and DNA methylation blocks only (the protein block does not contain any variable simulated with differential abundance, more details are given in Supplementary Materials section 5).

\begin{figure*}[ht]
    \centering
    \subfloat[Methylation]{
        \centering \includegraphics[width=0.47\textwidth, height=8cm]{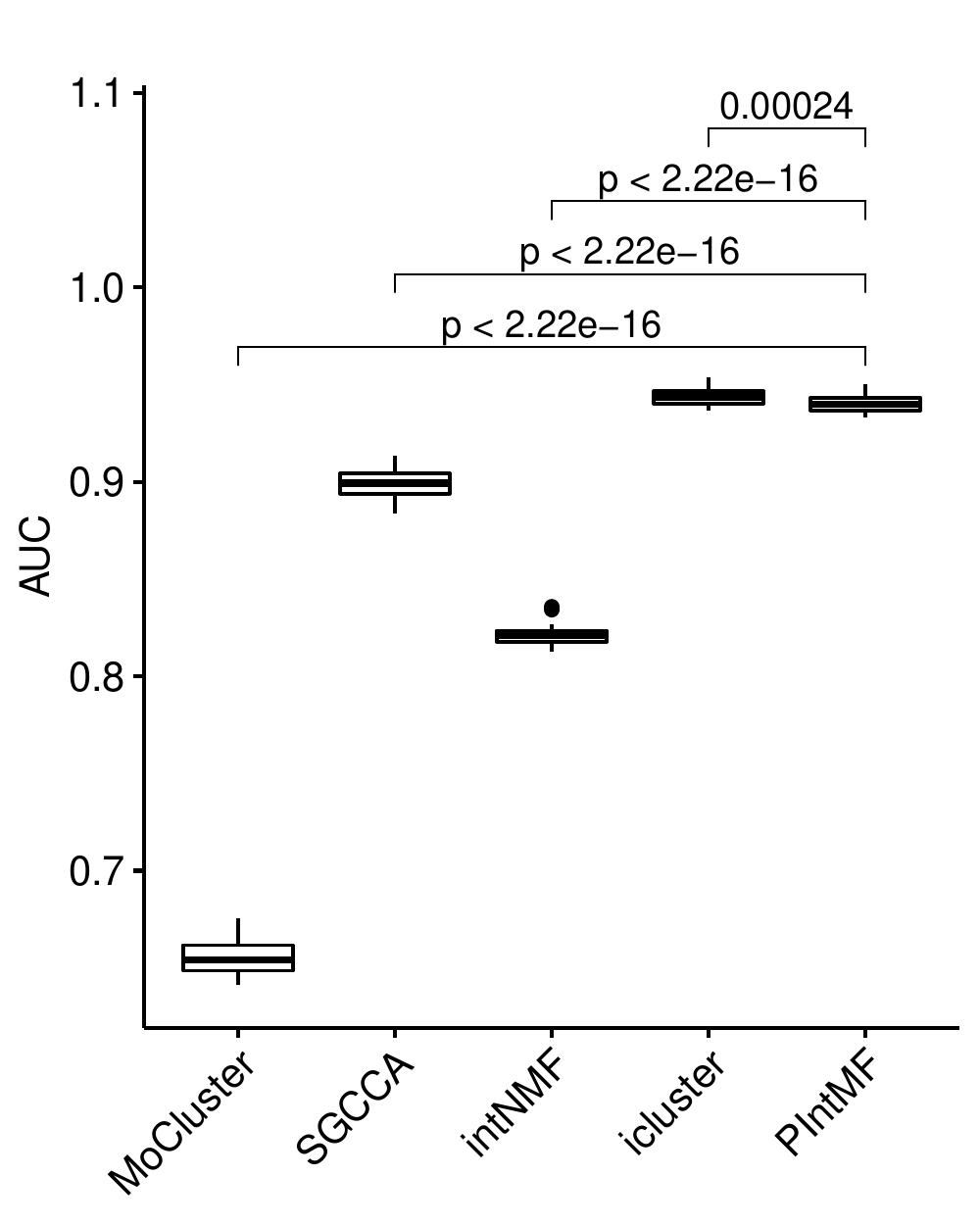}
        \label{fig:meth_omics_auc}
        }
\subfloat[Expression]{
        \centering \includegraphics[width=0.47\textwidth, height=8cm]{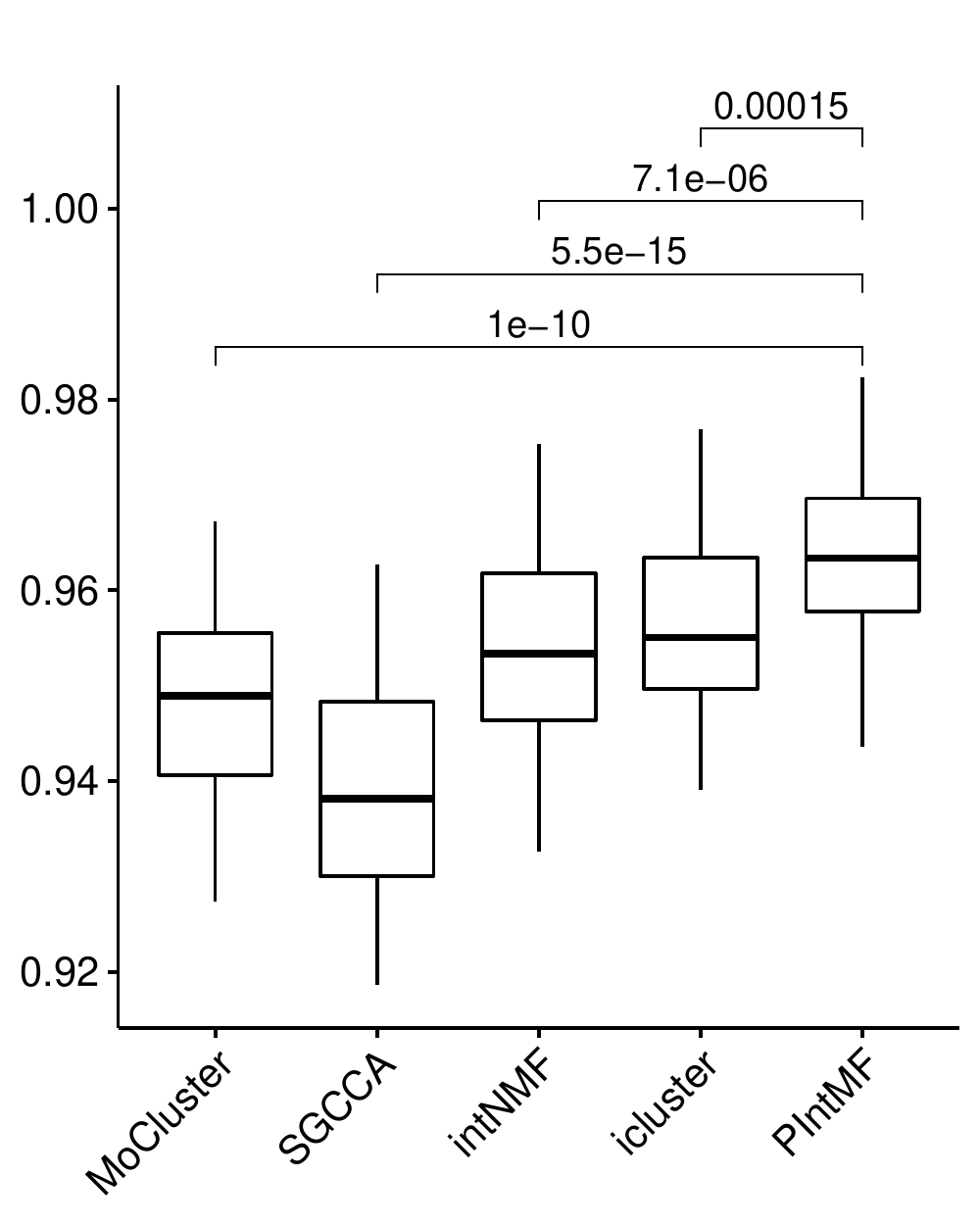}
        \label{fig:expr_omics_auc}
        }
        \caption{AUROC of PIntMF, MoCluster, SGCCA, iClusterPlus and intNMF for OmicsSIMLA simulations on (a) DNA methylation and  (b)  Gene expression blocks}
    \label{fig:auc-sim-omics-simla}
\end{figure*}

\textbf{DNA Methylation dataset:} PintMF and iclusterPlus outperform the others with similar performances but the AUROC of iclusterPlus is significantly higher. Then, the AUROC of PintMF is significantly higher than for MoCluster, SGCCA and intNMF (Fig. \ref{fig:auc-sim-omics-simla}).

\textbf{Expression dataset:} PIntMF is the best method with an AUROC significantly higher than the others. However, all methods achieve an AUROC higher than 0.92. (Fig. \ref{fig:auc-sim-omics-simla})

On these simulations, PIntMF gives similar results to iClusterPlus, but with automatic tuning of parameters. Besides, the algorithm of PIntMF is faster than iClusterPlus.

\subsubsection{Stability selection}
Jackknife was performed to evaluate the stability of variable selection. To perform this technique,  we run the model PIntMF on the data without one sample at each step. Therefore, we obtain $n$ datasets containing $n-1$ individuals on which we apply the method.

The stability of the selected variables for Binary, Gaussian, methylation and expression datasets seems to be strong (Fig. S10 in Supplementary Materials). For proteins and for beta-like data, the bootstrap reveals that some selected variables are not stable. The Jackknife method could be used to remove false-positives variables.

\subsubsection{Summary}

To summarize this simulation part (see Table \ref{tab:summary-perf}), our method PIntMF provides satisfying clustering and variable selection both on correlated blocks (Simulation Framework 2) and on non-correlated blocks (Simulation Framework 1). PIntMF is the only method that performs well on all simulated settings.

We conclude on these two frameworks of simulated data that PIntMF is a fast and flexible tool.

\begin{table}[ht!]
    \centering
    \begin{tabular}{lcccc}
                        &  \textbf{Clustering} & \shortstack{\textbf{Variable} \\\textbf{selection}} & \shortstack{\textbf{Automatic}\\ \textbf{Tunning}}& \shortstack{\textbf{Parameters} \\\textbf{left to tune}}\\
                        \hline
         iClusterPlus &+&++&-&>2\\
         intNMF &+++ &- &+++&1\\
         SGCCA &++&++&-&>5\\
         MoCluster&+++&+++&+&>2\\
         CIMLR&+&++&+++&1\\
                  PIntMF &+++&+++&+++&1\\

         \hline
    \end{tabular}
    \caption{Summary of the performance of the PIntMF compared to other methods}
    \label{tab:summary-perf}
\end{table}

\subsection{Applications}
\label{sec:real-dat}
In this section, we assess the performance of the PIntMF method on real data by considering two applications. The first one is a dataset from murine liver \citep{williams2016systems} under two different diets already used in two previous comparison articles \citep{pierre2019clustering, tini2017multi}, and the objective is to recover the diets of the mice (fat diet or chow diet). The second one is a glioblastoma dataset from TCGA used in \citep{shen2012integrative} and the goal is to find the tumor subtypes.

\subsubsection{PIntMF highlights variables linked to phenotypes of samples}

We analyzed the  BXD cohort (composed of 64 samples) \citep{williams2016systems}; the mice were shared into two different environmental conditions of diet: chow diet (CD) (6\% kcal of fat) or high-fat diet (HFD) (60\% kcal of fat). Measurements have been made in the livers of the entire population at the transcriptome, the proteome, and the metabolome levels. 

Therefore, we applied PIntMF to this dataset as well as intNMF, MoCluster, SGCCA, iClusterPlus, and CIMLR (Supplementary Materials Table S2).

PIntMF produces a perfect classification of the individuals for this real dataset .

For this dataset, all criteria for the model selection were computed (Supplementary Material Fig. S6), and 2 groups were selected for further analysis.

\begin{figure*}[ht]
    \centering
 \includegraphics[width=0.99\textwidth]{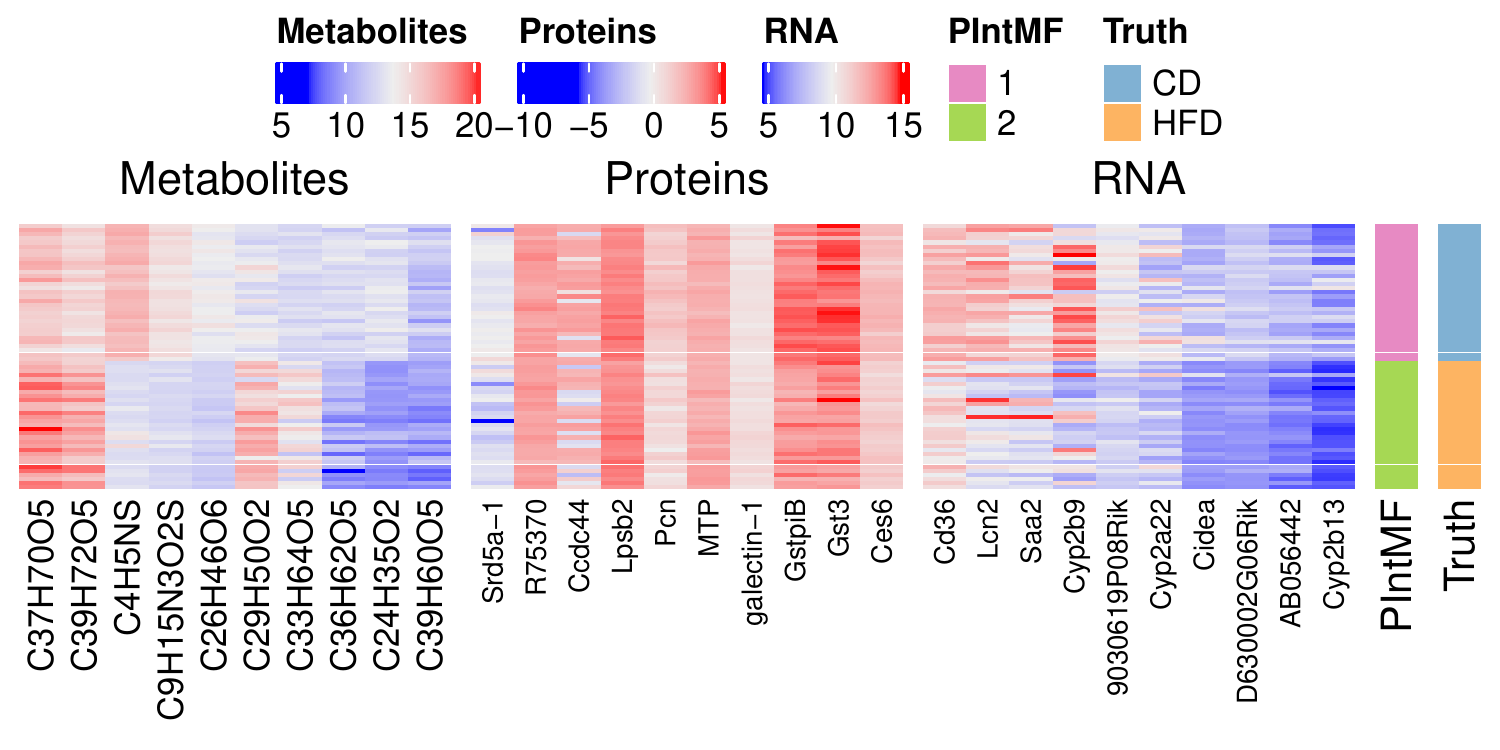}

  \caption{BXD cohort results: Top 10 selected variables with PIntMF of each dataset (Metabolites, Proteins and RNA), the clustering given by PIntMF and the true clustering are on the right.}
  \label{fig:heatmap-BXD}
\end{figure*}

PIntMF highlights interesting variables that seem to have different abundance between the two groups CD and HFD (Fig. \ref{fig:heatmap-BXD}): VITAMIN E (C$_29$H$_50$0$_2$), Cholesteryl (C$_36$H$_62$O$_5$), Mustard Oil (C$_4$H$_5$NS).
Saa2 gene that codes for a protein involved in the HDL complex seems to be deferentially expressed between the two groups. Then, the Cidea gene that is involved in the metabolism of lipids and lipoproteins has a slightly different level of expression between the two groups. Finally, Cyp2b9 oxidies steroids, fatty acids, and xenobiotics are less expressed in the high-fat diet group.
To conclude, PIntMF succeeds well to recover classification and relevant markers in all datasets.

\subsubsection{PIntMF reveals a new classification of non annotated samples on TCGA dataset.}

\begin{figure*}[ht]
    \centering
    \subfloat[Heatmap plot of $\bW$: Homogeneity between subtypes and subgroups identified by PIntMF ]{
        \centering \includegraphics[width=0.47\textwidth, height=8cm]{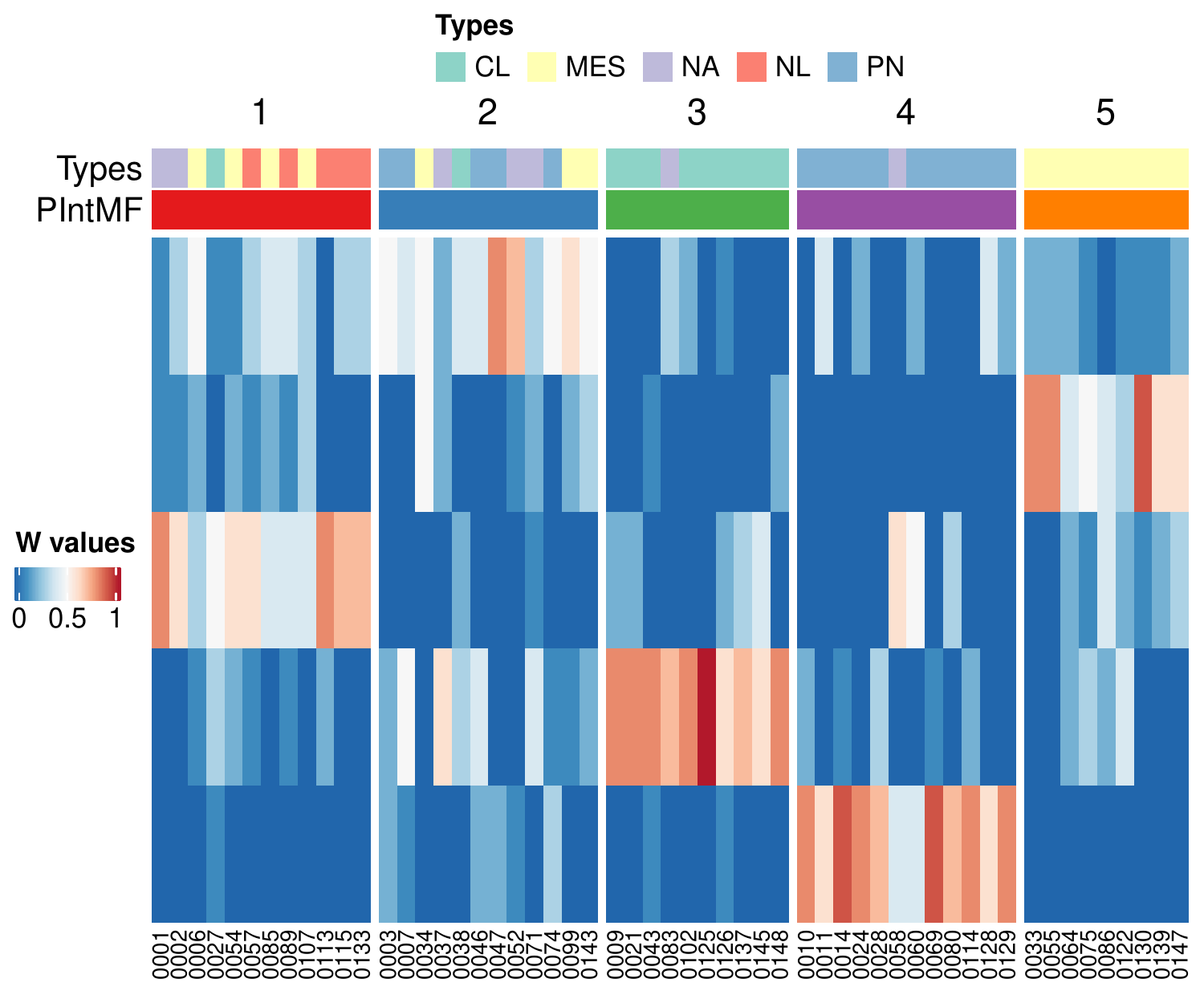}
        \label{fig:heatmap}
        }\quad
    \subfloat[Kaplan-Meier plot: The subgroups identified by PIntMF show survival differences]{
        \centering \includegraphics[width=0.47\textwidth,height=7.5cm]{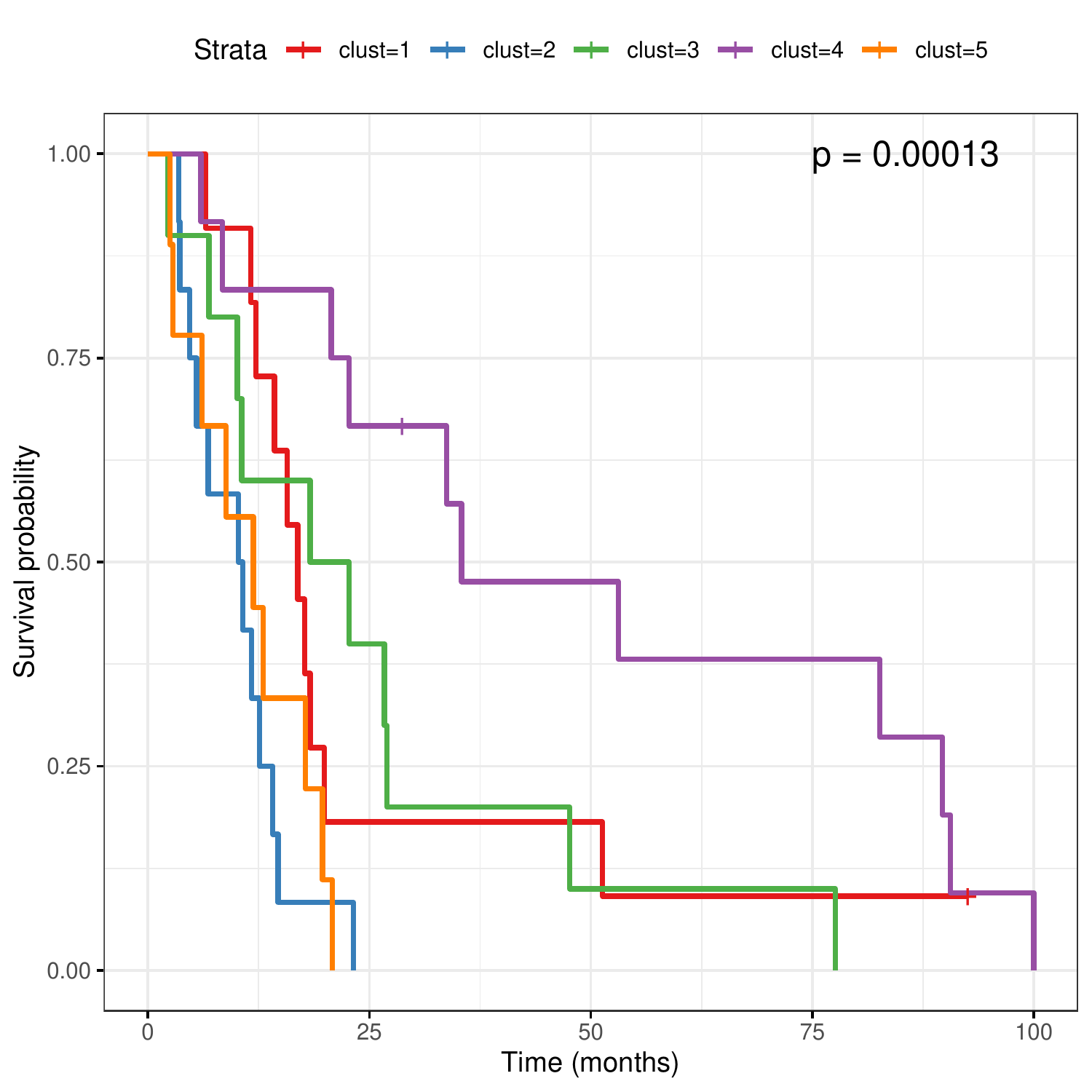}
        \label{fig:survival}
        }
        
    \subfloat[Copy number variation]{
        \centering \includegraphics[width=0.33\textwidth]{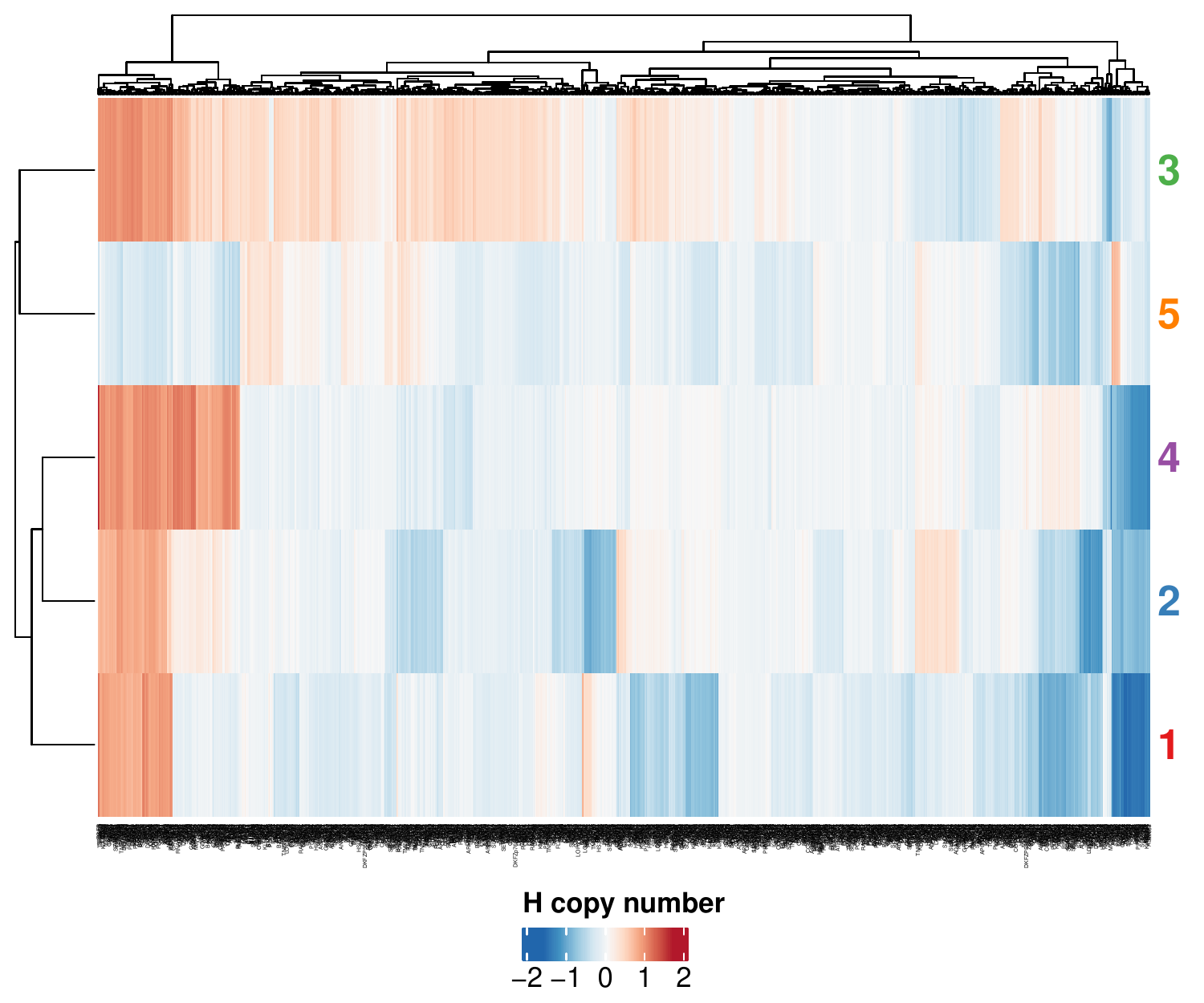}
        \label{fig:cn-gbm}
        }
    \subfloat[Gene Expression]{
        \centering \includegraphics[width=0.33\textwidth]{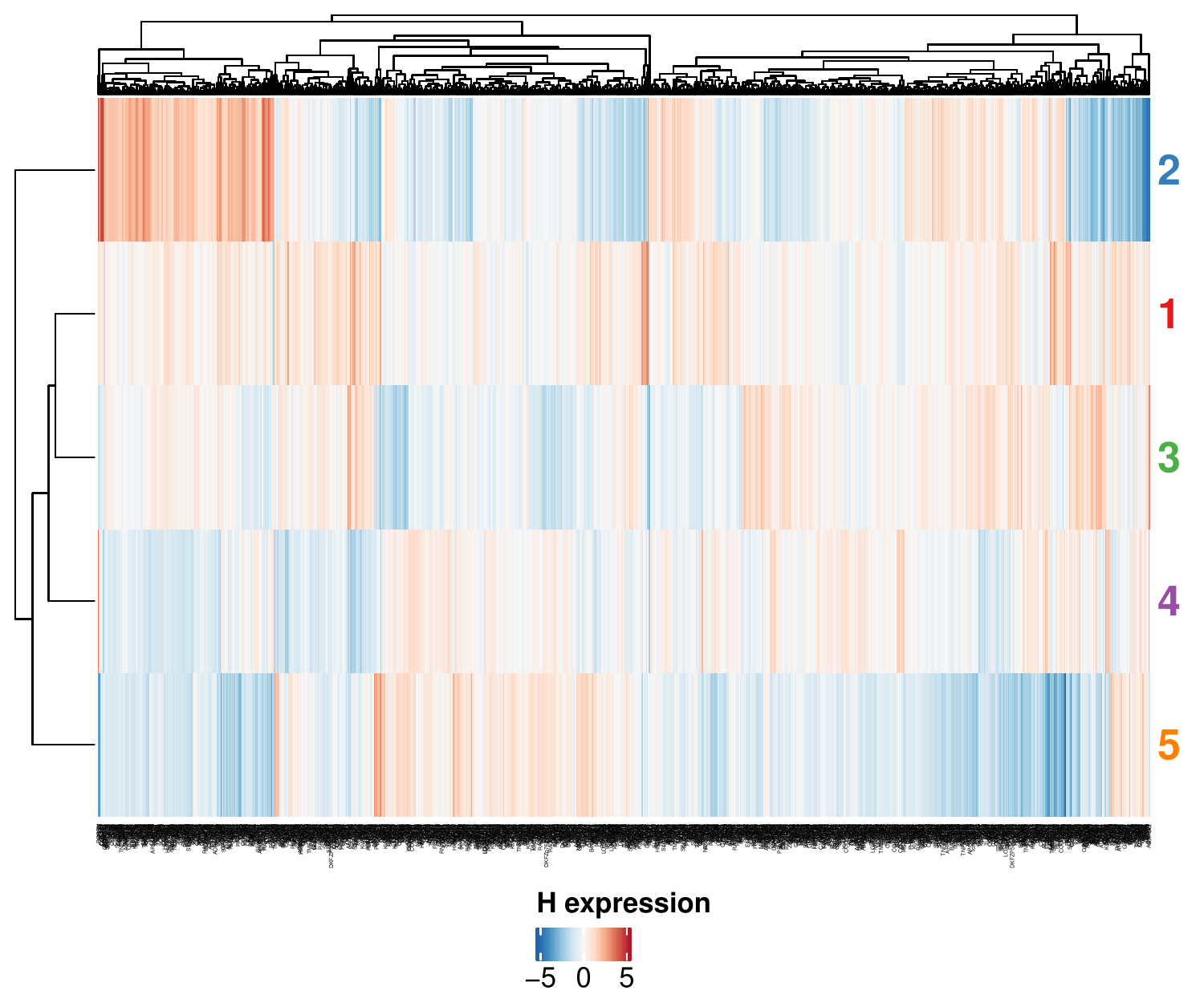}
        \label{fig:exp-gbm}
        }
        \subfloat[DNA Methylation]{
        \centering \includegraphics[width=0.33\textwidth]{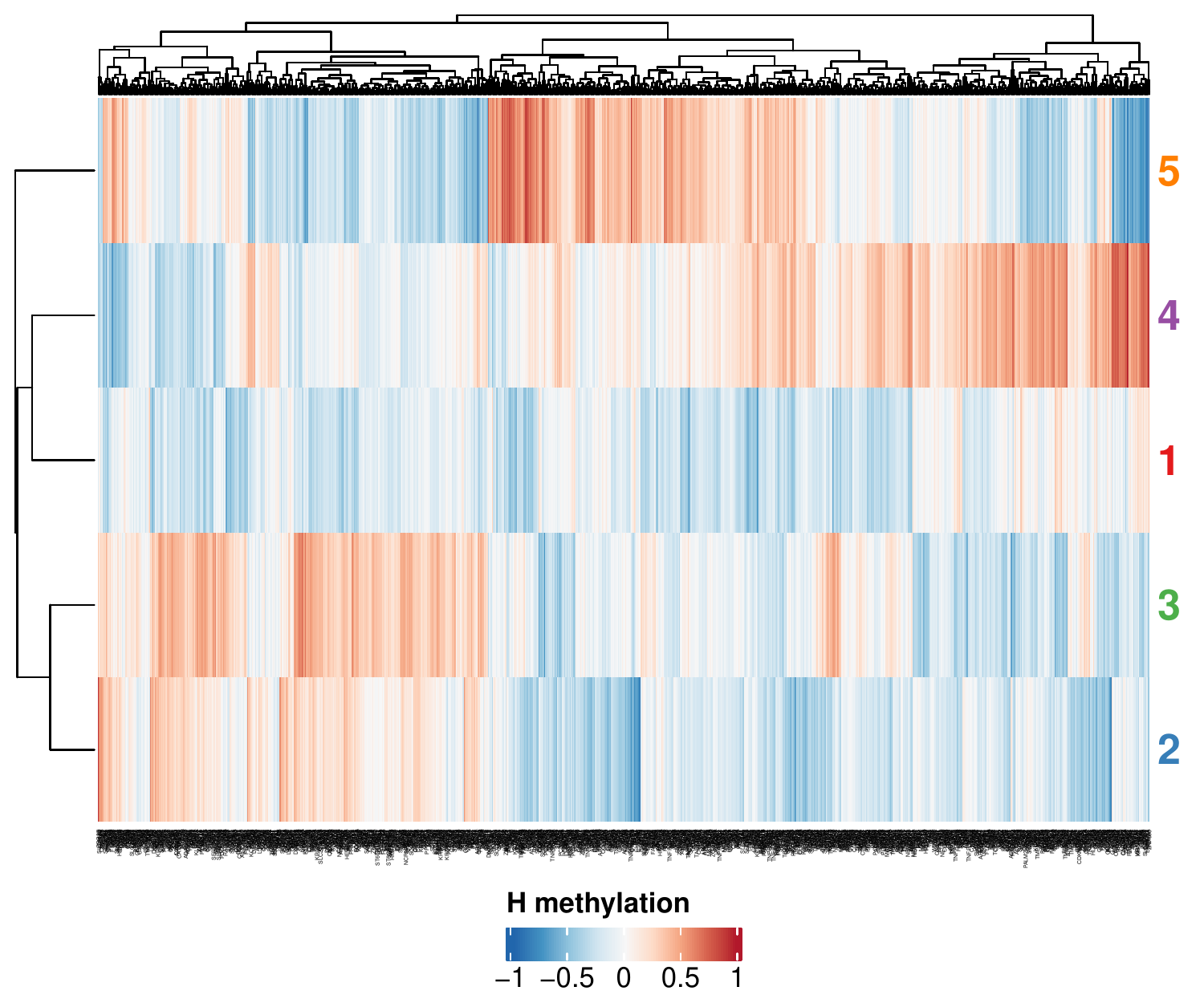}
        \label{fig:meth-gbm}
        }
        \caption{(a) Heatmap of $\bW$. The clustering of PIntMF was compared to glioblastoma subtypes. (b) Survival curves with p-value of log-rank test. (c, d, e)  $\bH$ matrix for the three considered omics blocs on glioblastoma dataset}
                \label{fig:gbm}
        \end{figure*}

Secondly, we analyze a subset of the glioblastoma dataset from the cancer genome atlas (TCGA): the Glioblastoma study (2009) used in \citep{shen2012integrative}. The dataset contains three matrices: copy number variation (1599 regions), DNA methylation (1515 CpG), and mRNA expression (1740 genes) in 55 samples. GBM samples were classified into four subtypes (Classical: CL, Mesenchymal: MES, Neural: NL, and Proneural: PN). Besides, there are samples with no subtype (NA). Using the PIntMF method, we highlight samples with no classification close to labeled samples.
Looking at the three criteria, the best number of latent variables seems to be 5 (Supplementary Materials Fig. S7).
For example, the green cluster from PIntMF matches a part of the CL subtype, and one sample labeled as NA is in this green cluster. 
Then, the purple cluster from PIntMF matches the PN subtype, and one sample labeled as NA can be classified with the PN subtype (Figure  \ref{fig:heatmap}).
Clusters 1 (red) and 2 (blue) are more heterogenous. However, the red one is composed of NL and NA labeled samples. The blue one is close to samples labeled as PN.

We performed a survival analysis to identify a relation between groups found by PIntMF and the survival rate (Figure  \ref{fig:survival}). The survival test gives a significant p-value at 5\% (p-value =0.00013 with log-rank test). The prognosis for the purple (4) group is better than those of the red and green (1 and 3) groups and even better than the orange and blue (2 and 5) groups. Note that the PN subtype is split into two groups (purple and blue) that have two very different survival curves.

The previous study \citep{shen2012integrative} performed with iCluster method \citep{shen2009integrative} identified 3 subgroups with a less significant p-value (0.01) than PIntMF for the survival differences between subgroups. Their Cluster 1 matches the PN group, Cluster 2 matches the CL group, and Cluster 3 is mostly composed of the MES subtype. Authors do not give any information about the samples with no subtypes.

$\bH$ matrices exhibit various types of genomic profiles according to the clusters (Figure \ref{fig:gbm}). For instance, the orange cluster (5) shows few alterations at the copy number variation level (Fig. \ref{fig:cn-gbm}) but a particular profile for DNA methylation and gene expression data (Fig. \ref{fig:meth-gbm}). The blue cluster (2) has a distinct pattern of expression (Fig. \ref{fig:exp-gbm}).

\section{Discussion}
We presented a new model to discover new subgroups of a cohort and potential new markers from several types of omic data. PIntMF is a matrix factorization model with positivity and sparsity constraints (Lasso) on inferred matrices. The method and all scripts of this article are available in an R package named PIntMF.

The main advantage of this method is the automatic tuning of the lasso penalties for both variable and sample matrices. To optimize the algorithm at the computational time level (Supplementary Materials Fig.S9), we tried several algorithms to infer matrices $\bH^k$.  \verb'glmnet' is very fast compared to the others (ncvreg, quadrupen, and biglasso), therefore it was retained for all analysis.  We also optimized the initialization of the algorithm that is obtained by using the SNF algorithm \citep{wang2014similarity}. This initialization provides, at the end of the algorithm, the best clustering and the best percentage of explained variation (Fig. S1). Besides, this initialization is performed at the integrative level rather than separately on each block of data. 

 PIntMF tunes automatically the penalties on matrices $\bH^k$ and $\bW$, without any intervention of the user, and we noticed that all the matrices are quite sparse on real datasets (Figure \ref{fig:gbm}). The user needs to choose only one parameter that is the number of latent variables. The last parameter can be chosen by looking at the MSE, cophenetic coefficient, and the PVE (Supplementary Materials Fig. S2 to S6). All these criteria are implemented in the R package. For non-correlated data simulations, only the cophenetic coefficient and the PVE allow choosing properly the correct number of latent variables.

It is still difficult to evaluate the performance of an integrative method on simulations \citep{Cantini2020}. The relationships between blocks of omics are complex, often not well-known, and the modeling of these links is not easy.  To our knowledge, there does not exist any reference dataset to assess performances. Therefore, we evaluated the algorithm on two different simulation frameworks (completely simulated and based on real-data) and two real datasets. Besides, we compared it with several other state-of-the-art integrative methods. 
We demonstrated, on the first simulated dataset (non-correlated blocks), that PIntMF outperforms the other methods on both clustering and variable selection. Indeed, on simulated data, the clustering from PIntMF makes few errors of classification. We also highlighted that PIntMF is more robust to heterogeneous data compared to the others: the method performs as well on gaussian distributions as on binary or beta distributions for the variable selection. 
On another simulated framework based on real data (correlated blocks), we observed good performances at clustering (perfect classification) and variable selection levels (AUROC upper than 90\%). With applications on two real datasets (BXD and TCGA data section \ref{sec:real-dat}), we demonstrated that the method could deal with real datasets. Besides, the application on the two real datasets shows that we found original subgroups but also interesting variables linked to the clinical phenotypes (diet and overall survival).

A weakness of the model is that the convergence of the algorithm to an optimal solution is not mathematically justified. Besides, a significance test for the variable selection is not given due to the use of the LASSO regression \citep{2021Jainlasso}. Jackknife could provide an idea of the confidence in the selected variables (Supplementary Materials Fig. S10). However, this type of approach is very time-consuming when datasets are large. 

Another improvement of the method could be dealing with missing values. Missing values could be inside a block for a few variables. These missing values could be imputed by the average of other correlated variables or by the values of the nearest neighbor or more complex methods as proposed by \citep{voillet2016handling,gonzalez2009highlighting,husson2013handling}. Commonly, a whole block can also be missing for an individual. In this case, the matrix $\bW$ could be computed only on the present blocks for this individual. Thanks to the $\bW$ matrix, we could deduce a new profile for this patient from the $\bH^k$ matrix inferred with the other individuals.

We could also extend PIntMF by including prior information such as the genome structure. For instance, we could force the algorithm to select the same genes in the DNA methylation block and the expression block. A group Lasso penalty \citep{simon2013sparse} could be added to the proposed model to include such a prior.

To conclude, PIntMF is an easy and flexible method to integrate omics data. It exhibits good performance in terms of classification or variable selection in both cases (correlated blocks or non-correlated blocks). Among all tested methods, it is the one that works in most situations. PIntMF is fast and automatically tunes the penalty for each block to select an appropriate number of variables (sparse matrices). Besides, it provides a sparse matrix $\bW$ to perform more easily the clustering of samples.
We also provide three criteria namely MSE, PVE, and cophenetic coefficient to choose the best number of latent variables.

The integration of several types of omics with our method could help in discovering potential markers even with a small number of patients. Finally, it could also help to classify patients with unknown phenotypes.

\section{Software}
An R package named PIntMF can be used to reproduce all simulations and figures and is available online at ??.

%

%

\bibliographystyle{natbib}
\bibliography{first.bib}

\end{document}


\maketitle

\section{Initialization of the algorithm}
\label{sec:init}
Often in NMF algorithms \citep{lee1999learning}, the matrices are initialized by non-negative random values.
Here, we evaluated four kinds of initialization for $\bW$ or $\bH$.

\textbf{SVD (Singular Value Decomposition)} 

SVD is also a matrix factorization technique with constraints. The SVD is defined as:
\begin{equation}
    SVD(\bX^k)= \mathbf{U}\mathbf{S}\mathbf{V}^T
\end{equation}
Where $\mathbf{U}, \mathbf{S}$ and $\mathbf{V}$ are of dimensions $n\times n$, $n\times J_k$ and $J_k\times J_k$. Matrix $\mathbf{S}$ is a diagonal matrix. If $r$ is the rank of matrix $\bX^k$, therefore $\mathbf{S}$ has $r$ non-zero entries. 

SVD provides the best low-rank linear approximation of the original matrix $\bX^k$ if we keep only $P\leq r$ singular values. The matrices $\mathbf{U}$ and $\mathbf{V}$  are also reduced to produce matrices $\mathbf{U}_P$  and
$\mathbf{V}_P$, respectively.  In our case, we initialize $\bH^k$ with $\mathbf{V}$.

\textbf{Hierarchical clustering} 

For this initialization, we perform a hierarchical clustering on each block and we keep the partitions in $P$ clusters. Then, average profiles from the clusters are computed.
For each block $k$:
\begin{enumerate}
    \item Compute Hierarchical clustering with Ward's method
    \item Cut the hierarchical clustering at $P$
    \item Obtain $P$ clusters denoted $\mathcal{C}_1, \dots, \mathcal{C}_P$
    \item $\bh^k_{p\bullet}= \frac{1}{Card(\mathcal{C}_p)}\sum_{i\in \mathcal{C}_p}\bx^k_{i\bullet}$
\end{enumerate}

\textbf{Random}

We sample $P$ profiles at random and $\bH^k$ are initialized with these profiles.

\textbf{SNF}

For this initialization, we initialize $\bW$ with the clustering in $P$ clusters from the algorithm SNF (Similarity network fusion \citep{wang2014similarity}). $\mathcal{C}_p$ denotes the cluster $p$ from SNF. Thus, 
\begin{equation}
    w_{ip} = \left\lbrace \begin{array}{cc}
         1 & \text{ if } i\in \mathcal{C}_p \\
          0 & \text{ otherwise}
    \end{array} \right.
\end{equation}

This initialization has the advantage to take into account simultaneously the $K$ blocks of the analysis.

\subsection{Best initialization method}

For non-negative matrix factorization methods, the first step consists in initializing either the matrix $\bW$ or in our case the matrices $\bH^k$.
Several kinds of initialization were compared: Hclust, SVD, random, and SNF (see section \ref{sec:init} for more details). To evaluate the influence on the final results of the type of initialization, 20 datasets were simulated with R package CriMMix previously developed by our group \cite{pierre2019clustering}. Each dataset is composed of three blocks (under Gaussian, Binary, Beta-Like distributions with respectively 100, 50 and 500 variables in each block). Four unbalanced groups have been simulated with respectively 5, 10, 20 and 25 individuals.  

First, ARIs between the clustering and the true clustering were computed for each type on initialization at the end of the algorithm (Fig. \ref{fig:ARI-init}). Second, we compute the PVE at each iteration to monitor the convergence of the algorithm (Fig. \ref{fig:PVE-init}).

The initialization with the SNF reaches higher values of both ARI and PVE. Indeed, ARI computed with SNF initialization is mainly close to 1 and PVE reaches fastly 20\%, for the other types of initialization a larger number of iterations is necessary to reach this level of PVE.
Hclust is the second best method to initialize the algorithm, however it does not use all the blocks jointly.
SVD is clearly not stable as well as random initialization.

For all the following analyses, SNF initialization was used.

\begin{figure*}[h]
    \centering
\subfloat[ARI]{
\includegraphics[width=0.49\textwidth]{Figs/eval_init.pdf}
\label{fig:ARI-init}
}
\subfloat[PVE]{
\includegraphics[width=0.49\textwidth]{Figs/pve_init.pdf}
\label{fig:PVE-init}
}
    \caption{Performance of the 4 tested types of initialization (hierachical clustering, Random, SNF and SVD). (a) Evaluation of the final clustering with adjusted Rand Index (ARI), and (b) Percentage of variation explained (PVE)}
    \label{fig:eval-init}
\end{figure*}

\section{Selection of the correct number of latent variables}

\subsection{The uncorrelated simulated dataset}
\label{sec:BIC}

For 25 simulations per benchmark, we ran the algorithm PIntMF for various numbers of latent variables (2 to 7). Then, we computed the different criteria (BIC, PVE and the cophenetic coefficient). We also added the RSS (Figures \ref{fig:bic-B1B2}, \ref{fig:bic-B3B4}, \ref{fig:bic-B56}, \ref{fig:bic-B7B8}).

On this dataset, the best criteria to choose the correct number of latent variables seem to be the cophenetic coefficient and the PVE.

For the benchmark 6 (with only two clusters), it seems that the criteria do not allow to choose the correct number of clusters. The model selection is difficult in the case where we simulated only two clusters in the data.

\begin{figure*}[h]
    \centering

\subfloat[Benchmark 1]{\includegraphics[width=0.90\textwidth]{Figs/Benchmark1_crit_test.pdf}}\\
\subfloat[Benchmark 2]{\includegraphics[width=0.90\textwidth]{Figs/Benchmark2_crit_test.pdf}}\\

 \caption{Benchmarks 1 and 2: Cophenetic coefficient, PVE, MSE for uncorrelated simulations. The vertical red line represents the simulated number of clusters.}
    \label{fig:bic-B1B2}
\end{figure*}

\begin{figure*}[h]
    \centering
\subfloat[Benchmark 3]{\includegraphics[width=0.90\textwidth]{Figs/Benchmark3_crit_test.pdf}}\\
\subfloat[Benchmark 4]{\includegraphics[width=0.90\textwidth]{Figs/Benchmark4_crit_test.pdf}}
\caption{Benchmarks 3 and 4: Cophenetic coefficient, PVE, MSE for uncorrelated simulations. The vertical red line represents the simulated number of clusters.}
    \label{fig:bic-B3B4}
\end{figure*}

\begin{figure*}[h]
    \centering
\subfloat[Benchmark 5]{\includegraphics[width=0.90\textwidth]{Figs/Benchmark5_crit_test.pdf}}\\
\subfloat[Benchmark 6]{\includegraphics[width=0.90\textwidth]{Figs/Benchmark6_crit_test.pdf}}\\
    \caption{Benchmarks 5 and 6: Cophenetic coefficient, PVE, MSE for uncorrelated simulations. The vertical red line represents the simulated number of clusters.}
    \label{fig:bic-B56}
\end{figure*}

\begin{figure*}[h]
    \centering

\subfloat[Benchmark7] {\includegraphics[width=0.90\textwidth]{Figs/Benchmark7_crit_test.pdf}}\\
\subfloat[Benchmark 8]{\includegraphics[width=0.90\textwidth]{Figs/Benchmark8_crit_test.pdf}}
    \caption{Benchmarks 7 and 8: Cophenetic coefficient, PVE, MSE for uncorrelated simulations. The vertical red line represents the simulated number of clusters.}
    \label{fig:bic-B7B8}
\end{figure*}

\subsection{Cophenetic coefficient, PVE, MSE on realistic simulations}
According to the three criteria, the best number of latent variables for PIntMF is 2. Indeed, this number minimizes the MSE criterion, adding more latent variables in the model does not really increase the PVE, and the cophenetic coefficient decreases just after 2.

\begin{figure*}[h]
    \centering
\includegraphics[width=0.90\textwidth]{Figs/OMICSSMILA_bics_pve.pdf}
    \caption{Cophenetic coefficient, PVE, MSE for correlated simulations}
    \label{fig:bic-per-omicsimla}
\end{figure*}

\subsection{Cophenetic coefficient, PVE, MSE on real datasets}

By looking at the MSE and PVE, we choose the best number of latent variables (P=5) for the Glioblastoma dataset (Fig. \ref{fig:GBM-bic-pve}). The cophenetic coefficient is not stable after p=6. 

By looking at the MSE, the best number of latent variable is 2 for the BXD dataset (Fig. \ref{fig:bxd-bic-pve}). The PVE slightly increases from p=2 to 5 latent variables, therefore we can consider that adding new latent variables is not useful. For the cophenetic coefficient, the best is for P=2 and 3.

\begin{figure}[ht]
    \centering
    \includegraphics[width=0.90\columnwidth]{Figs/GBM_bics_pve.pdf}
    \caption{Cophenetic coefficient, PVE, MSE for GBM data}
    \label{fig:GBM-bic-pve}
\end{figure}

\begin{figure}[ht]
    \centering
    \includegraphics[width=0.90\columnwidth]{Figs/BXD_bics_pve.pdf}
    \caption{Cophenetic coefficient, PVE, MSE for BXD data}
    \label{fig:bxd-bic-pve}
\end{figure}

\section{Fast computation for H matrix}
We optimized the computation of the $\bH$ matrix because it could be huge. Indeed, $\bH$ is the matrix that represents the variable profiles. The number of variables could range from one hundred to one billion.

Therefore, for the computation of the $\bH^k$ matrices, several packages \verb'R' have been tried (\verb'ncvreg' \citep{ncvreg}, \verb'glmnet' \citep{tibshirani1996regression}, \verb'biglasso' \citep{biglasso} and \verb'quadrupen' \citep{quadrupen}). All these packages proposed cross-validation (using 5 folds) to choose the optimal value of the lasso penalty.

\begin{figure}[h]
    \centering
\includegraphics[width=0.49\columnwidth]{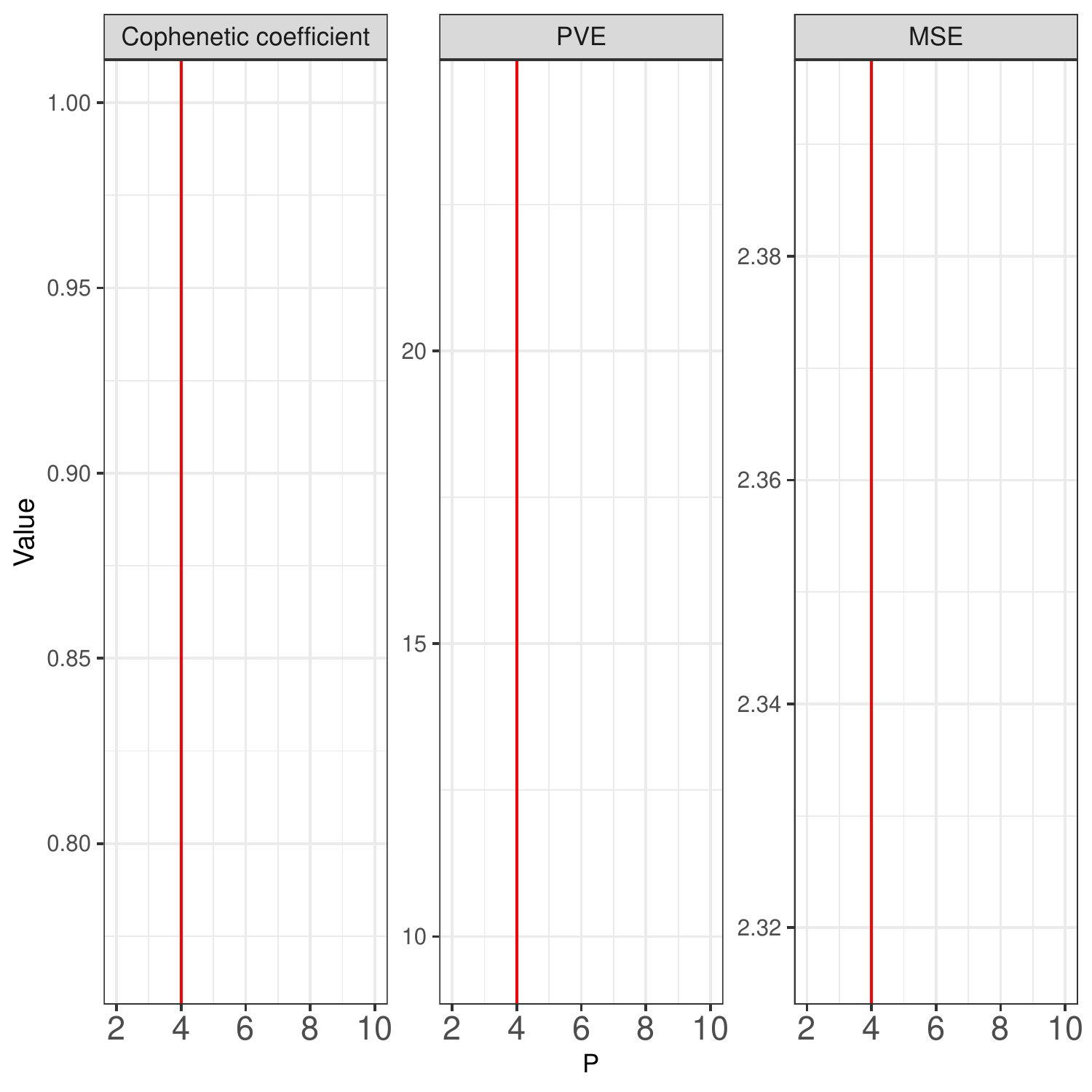}
\caption{Violin-plot of time computing of $\bH$ matrices for packages glmnet, quadrupen, bigmemory, and ncvreg.}
\label{fig:eval-lasso}
\end{figure}

The fastest method is clearly \verb'glmnet', therefore we used this package to solve $\bH^k$ matrices \ref{fig:eval-lasso}).

\section{Performance of PIntMF for variable selection}
Mean and standard deviation of AUROC were computed for PIntMF and the other 5 methods (Table \ref{tab:auc-sim}).

\begin{table*}[ht]
\centering
\begin{tabular}{llcccccc}
  \hline
 Benchmark & Dataset & CIMLR & iClusterPlus & iNMF & MoCluster & PIntMF & SGCCA \\ 
  \hline
\multirow{3}{*}{Benchmark1} & Beta-like & 0.93(0.01) & 0.84(0.04) & 0.29(0.01) & \textbf{0.97(0.01)} & 0.96(0.00) & 0.95(0.06) \\ 
   & Binary & 0.93(0.06) & 0.45(0.03) & 0.99(0.01) & \textbf{1.00(0.01)} & 0.98(0.02) & 0.95(0.06) \\ 
   & Gaussian & 0.99(0.02) & \textbf{1.00(0.01)} & \textbf{1.00(0.00)} &\textbf{1.00(0.00)} & \textbf{1.00(0.00)} & 0.98(0.05) \\ 
     \hline

  \multirow{3}{*}{Benchmark2} & Beta-like & 0.92(0.02) & 0.83(0.05) & 0.29(0.01) &\textbf{0.96(0.02)} &\textbf{ 0.96 (0.00)}& 0.94(0.006) \\ 
  & Binary & 0.92(0.05) & 0.46(0.03) & \textbf{0.99(0.01)}&\textbf{0.99(0.01)} & 0.98(0.02) & 0.93(0.07) \\ 
  & Gaussian & 0.97(0.03) & 0.96(0.04) & 0.99(0.01) & \textbf{1.00(0.00)} & \textbf{1.00(0.00)} & 0.96(0.06) \\
  \hline
\multirow{3}{*}{Benchmark3} & Beta-like & 0.94(0.01) & 0.85(0.04) & 0.29(0.01) &\textbf{ 0.98(0.01) }& 0.96(0.00) & 0.94(0.04) \\ 
  & Binary & 0.91(0.04) & 0.52(0.05) & 0.98(0.02) & \textbf{0.99(0.01)}& \textbf{0.99(0.02)}& 0.91(0.07) \\ 
  & Gaussian & \textbf{1.00(0.01)} & \textbf{1.00(0.00)} & \textbf{1.00(0.00)} & \textbf{1.00(0.00)} & \textbf{1.00(0.00)} & 0.97(0.06) \\ 
  \hline
\multirow{3}{*}{Benchmark4} & Beta-like & 0.87(0.01) & 0.69(0.05) & 0.32(0.01) & 0.87(0.01) &\textbf{ 0.88(0.01) }& 0.82(0.07) \\ 
  & Binary & 0.93(0.04) & 0.48(0.03) & 0.98(0.01) &\textbf{ 0.99(0.01) }& \textbf{0.99(0.02)} & 0.89(0.09) \\ 
 & Gaussian & 0.99(0.01) & 0.99(0.01) & \textbf{1.00(0.00)} & \textbf{1.00(0.00)} & \textbf{1.00(0.00)} & 0.92(0.09) \\ 
   \hline

 \multirow{3}{*}{Benchmark5} & Beta-like & 0.94(0.01) & 0.87(0.04) & 0.29(0.01) & \textbf{0.98(0.00)} & 0.96(0.00) & 0.97(0.00) \\ 
  & Binary & 0.95(0.02) & 0.45(0.02) & 0.99(0.01) & \textbf{1.00(0.01)} & 0.99(0.01) & 0.99(0.01) \\ 
  & Gaussian & \textbf{1.00(0.01)} & 0.99(0.02) & \textbf{1.00(0.00)} & \textbf{1.00(0.00)} & \textbf{1.00(0.00)} & \textbf{1.00(0.00)} \\ 
    \hline

\multirow{3}{*}{Benchmark6} & Beta-like & \textbf{1.00(0.00)} & 0.85(0.21) & 0.25(0.00) & 0.98(0.00) & \textbf{1.00(0.00)} & 0.97(0.01) \\ 
   & Binary & \textbf{1.00(0.00)} & 0.49(0.03) & \textbf{1.00(0.00)} & \textbf{1.00(0.00)} & \textbf{1.00(0.00)} & 0.92(0.03) \\ 
    & Gaussian & \textbf{1.00(0.01)} & 0.97(0.06) & \textbf{1.00(0.00)} & \textbf{1.00(0.00)} & \textbf{1.00(0.00)} & 0.99(0.04) \\  
    \hline
 \multirow{3}{*}{Benchmark7} & Beta-like & \textbf{1.00(0.00)} & 0.95(0.07) & 0.25(0.00) & \textbf{1.00(0.00)}& \textbf{1.00(0.00)} & \textbf{1.00(0.00)} \\ 
   & Binary &\textbf{1.00(0.00)} & 0.49(0.02) & \textbf{1.00(0.00)}& \textbf{1.00(0.00)}& \textbf{1.00(0.00)} & 0.98(0.02) \\ 
  & Gaussian & \textbf{1.00(0.00)}&\textbf{1.00(0.02)} &\textbf{1.00(0.00)} &\textbf{1.00(0.00)} & \textbf{1.00(0.00)} & \textbf{1.00(0.01)} \\ 
    \hline
 \multirow{3}{*}{Benchmark8} & Beta-like & \textbf{1.00(0.00)} & 0.98(0.04) & 0.26(0.00) & \textbf{1.00(0.00)} & \textbf{1.00(0.00)} &\textbf{1.00(0.00)} \\ 
  & Binary & \textbf{1.00(0.00)} & 0.47(0.03) & \textbf{1.00(0.00)} & \textbf{1.00(0.00)}& \textbf{1.00(0.00)} & 0.99(0.03) \\ 
  & Gaussian & \textbf{1.00(0.00)} & \textbf{1.00(0.00)} & \textbf{1.00(0.00)}& \textbf{1.00(0.00)}& \textbf{1.00(0.00)} & 0.99(0.03) \\ 
   \hline
\end{tabular}
\caption{Mean (standard deviation) of AUROC of intNMF, SGCCA, MoCluster, iClusterPlus, CIMLR and PIntMF methods on simulated data}
\label{tab:auc-sim}
\end{table*}

\section{Details on simulations OmicsSIMLA}
\textbf{Expression}
The tool OmicsSIMLA creates a file containing individual information, such as family ID, individual ID, father ID, mother ID, sex, and affection status. 
Then, the file containing the gene expression data (i.e., read counts) has various types of genes: EE (equivalently expressed genes between cases and controls), DE (differentially expressed genes between cases and controls), eQTL$\_$EE (equivalently expressed genes influenced by eQTL), eQTL$\_$DE (differentially expressed genes influenced by eQTL), eQTM$\_$EE (equivalently expressed genes influenced by eQTM), and eQTM$\_$DE (differentially expressed genes influenced by eQTM).
All columns labeled with DE correspond to the expression of the genes generated that are differentially expressed between case and controls.
Here, 100 genes have been generated with DE and 3 genes are influenced by the eQTM.

\textbf{Methylation}
The file containing methylation data is composed of the methylated read counts and total read depth for each CpG (separated by comma).
We computed the percentage of methylation values from this table with an home-made R script. A file indicates which probes are DM (differentially methylated) and the eQTM status is also generated.

For the five methods, we transform the beta values to M-values by using this formula:
 $$M_i = log2((\beta_i+\epsilon)/(1-(\beta_i+\epsilon))$$
Note that, some beta values are exactly one or zero. To avoid getting infinite values for $M$, the same value $\epsilon=0.001$ is added for each beta values before applying the formula.

\textbf{Proteomics}
They used the mass-action kinetic action model \citep{guoshouteo2015mass} to simulate proteomics data at a certain time point incorporating the gene expression data. No proteins are simulated as "true positive" variables to drive the clusters. Therefore, we do not evaluate the performance of the methods in terms of variable selection on this block.

Finally, the dataset is composed of 100 individuals and three omics blocks with 12102 genes with expression data, 2248 CpG probes, and 12103 proteins respectively.

\section{Jackknife}
Jackknife was performed to evaluate the stability of variable selection. To perform this technique,  we run the model PIntMF on the data without one individual at each step. Therefore, we obtain $n$ datasets containing $n-1$ individuals on which we apply the method.

\begin{figure*}[h]
    \centering
\subfloat[Correlated simulated datasets]{
\includegraphics[width=0.49\textwidth]{Figs/bootstrap_omicssimulation.pdf}
\label{fig:boot_omics}
}
\subfloat[Non-correlated simulated datasets]{
\includegraphics[width=0.49\textwidth]{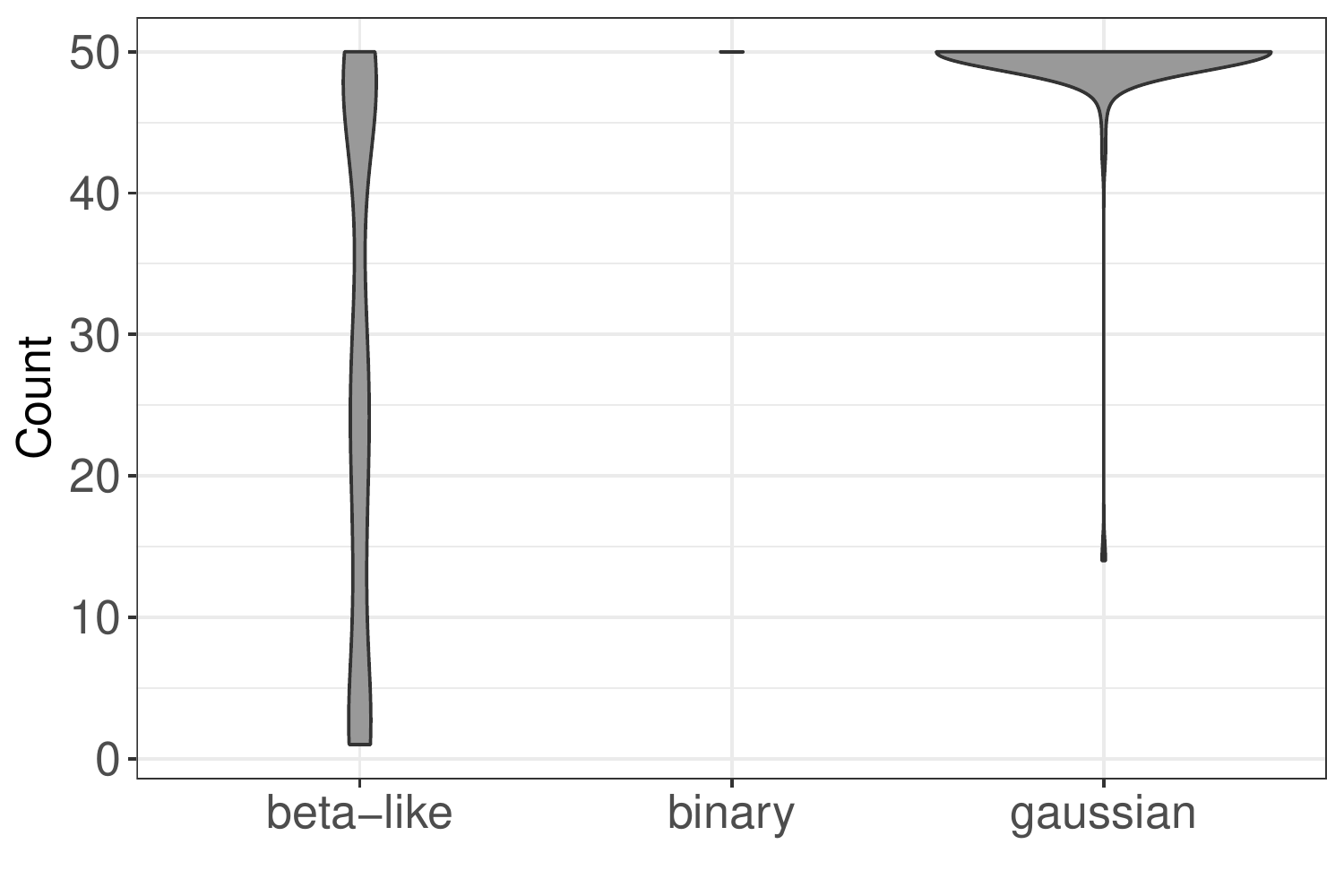}
\label{fig:boot_crimmix}
}
    \caption{Jackknife on simulated datasets. At each run, one sample is removed, we count the number of times where each variable is selected by the model (coefficient not equal to zero).}
    \label{fig:boot-sim}
\end{figure*}

The stability of the selected variables for Binary, Gaussian, methylation and expression datasets seems to be robust (Figures \ref{fig:boot_omics} and \ref{fig:boot_crimmix}). For proteins and for beta-like data, the bootstrap reveals that some selected variables are not stable and we could remove false-positives by adding this step.

\section{Additional results on BXD dataset}
\subsection{Clustering}

\begin{table}[ht]
\centering

\begin{tabular}{cllr}
  \hline
  \textbf{Data Set} &  \textbf{Methods} &  \textbf{F-score} &  \textbf{ARI} \\
  \hline
\multirow{8}{*}{BXD} & iClusterPlus & 0.97 & 0.88 \\ 
  & MoCluster & 1.00 & 1.00\\
  & intNMF & 0.98 & 0.94\\
  & SGCCA & 0.95 & 0.82\\
  & CIMLR & 1.00 & 1.00 \\
  & PIntMF & 1.00 & 1.00 \\
  \hline
\end{tabular}
\caption{Performance evaluation of clustering  using ARI and F-measure on BXD data.}
\label{tab:real-ari-fmeas}
\end{table}

\section{Additional results on Glioblastoma dataset}

\subsection{Comparison of selected variables}
We explore the selected variables on each dataset (Copy number, expression, and methylation). We notice that most selected variables are specific to one component. For instance, for the copy number dataset and the 3rd component of the model, 136 variables are specific to this component. We conclude that these variables are specific to the corresponding cluster and could have an impact on survival. Some variables are also shared between several components (Fig. \ref{fig:upset-gbm}) and could reveal common biological processes.

\begin{figure*}[h]
    \centering
\subfloat[Copy Number variables]{
\includegraphics[width=0.80\textwidth]{Figs/cn_upset_gbm.pdf}
\label{fig:cn-upset}
}\\
\subfloat[Expression variables]{
\includegraphics[width=0.80\textwidth]{Figs/exp_upset_gbm.pdf}
\label{fig:exp-upset}
}\\
\subfloat[Methylation variables]{
\includegraphics[width=0.80\textwidth]{Figs/methylation_upset_gbm.pdf}
\label{fig:meth-upset}
}
    \caption{GBM: Comparison of selected variables across the latent variables from PIntMF for each dataset.}
    \label{fig:upset-gbm}
\end{figure*}

\subsection{Enrichment}
We performed a pathway enrichment analysis with the R package \texttt{ClusterProfiler} \citep{clusterProfiler} for each component with the union of the selected genes across the three datasets. As the selected genes, some pathways are unique to the components (Fig. \ref{fig:enrich_gbm}), this means that some pathways could be linked to survival.
All pathways for each component are in table \ref{tab:gbm:component_pathways}.

\begin{figure*}
    \centering
    \includegraphics[width=0.89\textwidth]{Figs/upset_enrichment_GBM.pdf}
    \caption{GBM: Comparison of enriched  pathways across latent variables.}
    \label{fig:enrich_gbm}
\end{figure*}

\begin{table}[ht]
\centering
\begin{tabular}{rlrr}
  \hline
 & Description & p\_value & Comp \\ 
  \hline
1 & Proteoglycans in cancer & 0.00 &   1 \\ 
  2 & Focal adhesion & 0.00 &   1 \\ 
  3 & Transcriptional misregulation in cancer & 0.00 &   1 \\ 
  4 & Leishmaniasis & 0.00 &   1 \\ 
  5 & Chagas disease & 0.00 &   1 \\ 
  6 & Amoebiasis & 0.00 &   1 \\ 
  7 & Legionellosis & 0.00 &   1 \\ 
  8 & Rheumatoid arthritis & 0.00 &   1 \\
  \hline
  9 & Focal adhesion & 0.00 &   2 \\ 
  10 & AGE-RAGE signaling pathway in diabetic complications & 0.00 &   2 \\ 
  11 & Amoebiasis & 0.00 &   2 \\ 
  12 & PI3K-Akt signaling pathway & 0.00 &   2 \\ 
  13 & Pertussis & 0.00 &   2 \\ 
  14 & Cell adhesion molecules & 0.00 &   2 \\ 
  15 & Legionellosis & 0.00 &   2 \\ 
  16 & Transcriptional misregulation in cancer & 0.00 &   2 \\ 
  17 & Rheumatoid arthritis & 0.00 &   2 \\ 
  18 & Staphylococcus aureus infection & 0.00 &   2 \\ 
  19 & Hematopoietic cell lineage & 0.00 &   2 \\ 
  20 & PPAR signaling pathway & 0.00 &   2 \\ 
  21 & TNF signaling pathway & 0.00 &   2 \\ 
  22 & ECM-receptor interaction & 0.00 &   2 \\ 
  23 & NF-kappa B signaling pathway & 0.00 &   2 \\ 
  \hline
  24 & p53 signaling pathway & 0.00 &   3 \\ 
  25 & Cell cycle & 0.00 &   3 \\ 
  26 & HIF-1 signaling pathway & 0.00 &   3 \\ 
  27 & Bile secretion & 0.00 &   3 \\ 
  28 & DNA replication & 0.00 &   3 \\ 
  29 & Fc gamma R-mediated phagocytosis & 0.00 &   3 \\ 
  30 & Mineral absorption & 0.00 &   3 \\ 
  31 & Rheumatoid arthritis & 0.00 &   3 \\ 
  \hline
  32 & Cell adhesion molecules & 0.00 &   4 \\ 
  33 & Proteoglycans in cancer & 0.00 &   4 \\ 
  34 & Rheumatoid arthritis & 0.00 &   4 \\ 
  35 & HIF-1 signaling pathway & 0.00 &   4 \\ 
  36 & Focal adhesion & 0.00 &   4 \\ 
  37 & Ras signaling pathway & 0.00 &   4 \\ 
  38 & Transcriptional misregulation in cancer & 0.00 &   4 \\ 
  \hline
  39 & Phospholipase D signaling pathway & 0.00 &   5 \\ 
  40 & AGE-RAGE signaling pathway in diabetic complications & 0.00 &   5 \\ 
  41 & Chagas disease & 0.00 &   5 \\ 
  42 & Focal adhesion & 0.00 &   5 \\ 
  43 & Complement and coagulation cascades & 0.00 &   5 \\ 
  44 & Glutamatergic synapse & 0.00 &   5 \\ 
   \hline
\end{tabular}
\caption{Enriched pathway for each component from PIntMF}
\label{tab:gbm:component_pathways}
\end{table}

\clearpage
\normalsize

\bibliographystyle{natbib}
\bibliography{first}